\documentclass[fleqn,usenatbib]{mnras}

\usepackage{newtxtext,newtxmath}
\usepackage[T1]{fontenc}
\usepackage{ae,aecompl}

\usepackage{graphicx}	% Including figure files
\usepackage{amsmath}	% Advanced maths commands
\usepackage{multicol} 
\usepackage{multirow}
\usepackage{comment}
\usepackage{array}
\usepackage{soul}
\hypersetup{colorlinks=true,citecolor=blue,linkcolor=purple,filecolor=black,runcolor=black,breaklinks=true}
\usepackage{etoolbox}
\usepackage{ae,aecompl}
\usepackage[usenames]{color}
\usepackage{hyperref}
\usepackage{threeparttable}
\usepackage{mathrsfs}
\usepackage{scalerel}
\usepackage{longtable,booktabs,threeparttablex}
\usepackage{float}
\usepackage{stfloats}

\definecolor{purple}{RGB}{160,32,240}

\definecolor{red}{RGB}{225,50,50}

\definecolor{addchange}{RGB}{215,25,25}

\definecolor{removechange}{RGB}{25,25,215}

\newcommand{\HST}{\emph{HST}}
\newcommand{\JWST}{\emph{JWST}}
\newcommand{\Spitzer}{\emph{Spitzer}}
\newcommand{\Muv}{\ensuremath{\mathrm{M}_{\mathrm{UV}}^{ }}}

\newcommand{\logten}{\ensuremath{\log_{10}}}

\newcommand{\Lya}{\ensuremath{\mathrm{Ly}\alpha}}
\newcommand{\zLya}{\ensuremath{z_{_{\mathrm{Ly\alpha}}}}}

\newcommand{\xHI}{\ensuremath{x_{\mathrm{HI}}}}

\newcommand{\xiion}{\ensuremath{\xi_{\mathrm{ion}}}}
\newcommand{\logxiion}{\ensuremath{\log_{10}[\xi_{\mathrm{ion}} / (\mathrm{erg}^{-1}\ \mathrm{Hz})]}}
\newcommand{\fesc}{\ensuremath{f_{\mathrm{esc}}}}
\newcommand{\OIIIHb}{[OIII]+H\ensuremath{\beta}}

\newcommand{\Msol}{\ensuremath{\mathrm{M}_{\odot}}}
\newcommand{\Mstar}{\ensuremath{\mathrm{M}_{\ast}}}

\newcolumntype{P}[1]{>{\centering\arraybackslash}p{#1}}
\newcommand\Tstrut{\rule{0pt}{2.6ex}}         % = `top' strut
\newcommand\Bstrut{\rule[-1.2ex]{0pt}{0pt}}   % = `bottom' strut

\title[A Very Large Ionized Bubble in COSMOS?]{Strong Ly\boldmath{$\alpha$} Emission in an Overdense Region at \boldmath{$z=6.8$}: A Very Large (\boldmath{$R\sim3$} physical Mpc) Ionized Bubble in COSMOS?}

\author[R. Endsley \& D.P. Stark]{
Ryan Endsley$^{1}$\thanks{E-mail: rendsley@email.arizona.edu} \&
Daniel P. Stark$^{1}$
\\
$^{1}$Steward Observatory, University of Arizona, 933 N Cherry Ave, Tucson, AZ 85721 USA\\
}

\date{Accepted XXX. Received YYY; in original form ZZZ}

\pubyear{2022}

\begin{document}
\label{firstpage}
\pagerange{\pageref{firstpage}--\pageref{lastpage}}
\maketitle


\begin{abstract}
Our understanding of reionization has advanced considerably over the past decade, with several results now demonstrating that the IGM transitioned from substantially neutral at $z=7$ to largely reionized at $z=6$.
However, little remains known about the sizes of ionized bubbles at $z\gtrsim7$ as well as the galaxy overdensities which drive their growth.
Fortunately, rest-UV spectroscopic observations offer a pathway towards characterizing these ionized bubbles thanks to the resonant nature of Lyman-alpha photons.
In a previous work, we presented Ly$\alpha$ detections from three closely-separated Lyman-break galaxies at $z\simeq6.8$, suggesting the presence of a large ($R>1$ physical Mpc) ionized bubble in the 1.5 deg$^2$ COSMOS field.
Here, we present new deep Ly$\alpha$ spectra of ten UV-bright ($\mathrm{M}_{\mathrm{UV}}^{} \leq -20.4$) $z\simeq6.6-6.9$ galaxies in the surrounding area, enabling us to better characterize this potential ionized bubble.
We confidently detect (S/N$>$7) Ly$\alpha$ emission at $z=6.701-6.882$ in nine of ten observed galaxies, revealing that the large-scale volume spanned by these sources (characteristic radius $R=3.2$ physical Mpc) traces a strong galaxy overdensity ($N/\langle N\rangle \gtrsim 3$).
Our data additionally confirm that the Ly$\alpha$ emission of UV-bright galaxies in this volume is significantly enhanced, with 40\% (4/10) showing strong Ly$\alpha$ emission (equivalent width$>$25 \AA{}) compared to the 8--9\% found on average at $z\sim7$.
The median Ly$\alpha$ equivalent width of our observed galaxies is also $\approx$2$\times$ that typical at $z\sim7$, consistent with expectations if a very large ($R\sim3$ physical Mpc) ionized bubble is allowing the Ly$\alpha$ photons to cosmologically redshift far into the damping wing before encountering HI.
\end{abstract}

%%%%%%%%%%%%%%%%%%%%%%%%%%%%%%%%%%%%%%%%%%%%%%%%%%

%%%%%%%%%%%%%%%%% BODY OF PAPER %%%%%%%%%%%%%%%%%%

\defcitealias{Endsley2021_OIII}{E21a}
\defcitealias{Endsley2021_LyA}{E21b}

\section{Introduction} \label{sec:intro}

The epoch of reionization marks when radiation produced within the first galaxies ionized almost every hydrogen atom in the intergalactic medium (IGM).
Over the past two decades, much effort has been devoted to understanding this connection between galaxy formation and the large-scale ionization state of the early Universe \citep{Robertson2021}.
One of the primary observational tools used to study the timeline of reionization is the Lyman-alpha emission line from high-redshift galaxies.
Due to its resonant nature, \Lya{} is highly susceptible to scattering by HI, and thus will weaken considerably at epochs when the IGM is significantly neutral \citep[e.g.][]{MiraldaEscude1998,Haiman2002,Mesinger2004,McQuinn2007_LyA}.
Many spectroscopic surveys have revealed that the fraction of typical ($\Muv{} \gtrsim -20$) star-forming galaxies showing strong \Lya{} falls abruptly at $z>6$, suggesting that the IGM transitioned from largely ionized to significantly neutral between $z=6$ and $z=7$ \citep{Fontana2010,Stark2010,Ono2012,Caruana2014,Pentericci2014,Pentericci2018,
Schenker2014,Tilvi2014,Jung2017,Jung2020,Mason2018_IGMneutralFrac,Mason2019a,
Hoag2019,Fuller2020,Whitler2020}.
Such a reionization timeline is consistent with spectra of high-redshift quasars which indicate a low IGM neutral fraction at $z\simeq6$ ($\xHI{} \lesssim 10$\%; \citealt{McGreer2015}), yet a substantially neutral IGM at $z\simeq7-7.5$ ($\xHI{} \sim 50$\%; \citealt{Davies2018,Wang2020,Yang2020_Poniuaena}).

A key next step in studying reionization is understanding the growth of intergalactic HII regions.
Due to spatial variations in structure formation, reionization is predicted to have been a patchy process where ionized bubbles covered certain portions of the early Universe while other volumes remained highly neutral \citep[e.g.][]{Ciardi2000,MiraldaEscude2000}.
The first large ionized bubbles are expected to have formed around galaxy overdensities given the excess number of ionizing photons produced with them \citep[e.g.][]{Barkana2004,Furlanetto2004,Iliev2006,Castellano2016,Hutter2017,Kannan2021,Leonova2021,Garaldi2022}.
Over the coming decade, much effort is aimed at characterizing the size, spatial distribution, and associated overdensities of these ionized bubbles to provide detailed insight into how the IGM reionized.

One pathway towards identifying and mapping large ionized structures in the early IGM is using \Lya{} observations \citep[e.g.][]{McQuinn2007_LyA,Jensen2014,Castellano2018,Tilvi2020,Jung2020,Endsley2021_LyA,Hu2021}.
When \Lya{} photons are emitted from galaxies within large ionized bubbles, they cosmologically redshift far into the damping wing before encountering neutral hydrogen, and hence transmit more efficiently through the IGM \citep{Wyithe2005,Furlanetto2006,Weinberger2018,Mason2020,Park2021,Smith2021,Qin2022}.
Accordingly, a key indicator of a large ionized bubble is an enhanced \Lya{} equivalent width (EW) distribution among galaxies within that volume.
While it is straightforward to test if certain regions of the early Universe show enhanced \Lya{}, it has long been challenging to survey a big enough area on the sky to statistically sample large intergalactic HII regions.
For reference, the largest ionized bubbles at $z=7$ are predicted to span $\approx$15--30 arcmin ($\approx$5--10 physical Mpc) in diameter \citep{Lin2016}, suggesting that the identification of even just one of these bubbles may require \Lya{} observations covering multiple square degrees.
A compounding challenge is that these wide areas must contain deep (m$\gtrsim$26) near-infrared imaging since a rest-UV continuum measurement is necessary to determine the \Lya{} EW of individual galaxies.

In recent years, multiple ground-based imaging campaigns have begun reaching depths necessary to identify hundreds of $z\gtrsim7$ Lyman-break galaxies across several square degrees \citep[e.g.][]{Bowler2014,Bowler2020,Stefanon2019,Endsley2021_OIII,Harikane2021}.
Spectroscopic follow-up of these sources is now underway and initial observations have already delivered a handful of \Lya{} detections out to $z=7.2$ \citep{Ono2012,Furusawa2016,Bouwens2021_REBELS,Endsley2021_LyA}.
One notable outcome of these first results was the detection of \Lya{} from three closely-separated UV-bright ($\Muv{} \leq -20.4$) galaxies at $z\simeq6.8$ situated in the 1.5 deg$^2$ COSMOS field (\citealt{Endsley2021_LyA}, hereafter \citetalias{Endsley2021_LyA}).
Because \Lya{} is rarely seen from reionization-era galaxies, it was suggested that these sources may trace a large ionized bubble.
The surrounding field also showed a high surface density of $z\simeq6.8$ Lyman-break galaxies, hinting at the presence of a strong galaxy overdensity which would be expected to help generate a large intergalactic HII region.

In this paper, we present deep \Lya{} spectroscopy of ten UV-bright ($\Muv{} \leq -20.4$) $z\simeq6.8$ galaxies around the tentative ionized bubble identified in \citetalias{Endsley2021_LyA}.
We detect \Lya{} emission from nine of ten targeted galaxies (\S\ref{sec:results}), confirming that the volume spanned by these sources (characteristic radius $R=3.2$ physical Mpc) traces a strong overdensity ($N/\langle N\rangle \gtrsim 3$; \S\ref{sec:overdensity}).
Our data also verify that the \Lya{} emission of UV-bright galaxies in this volume is significantly enhanced relative to average (\S\ref{sec:bubble}), consistent with a picture wherein the surrounding IGM is highly reionized.

Throughout this paper, we quote magnitudes in the AB system \citep{OkeGunn1983}, employ a \citet{Chabrier2003} stellar initial mass function (IMF), and report 68\% confidence interval uncertainties.
We also quote distances in physical units and adopt a flat $\Lambda$CDM cosmology with parameters $h = 0.7$, $\Omega_\mathrm{M} = 0.3$, and $\Omega_\mathrm{\Lambda} = 0.7$.

\begin{table*}
\centering
\caption{Positions and photometric properties of the twelve $z\simeq6.6-6.9$ galaxies identified across an 11$\times$15 arcmin$^2$ region in COSMOS surrounding the tentative ionized bubble reported in \citetalias{Endsley2021_LyA}. For sources with a non-detection (S/N$<$1) in one of the IRAC bands, we report the 2$\sigma$ limiting color. We do not report the IRAC color for COS-957862 given the strong confusion from a bright neighboring source. In the final column we report the total MMT/Binospec exposure time and co-added average seeing for each targeted source. Neither COS-957862 nor COS-1038989 have yet been observed with Binospec.} 
\label{tab:table1}
\begin{tabular}{P{1.8cm}P{1.4cm}P{1.5cm}P{1.3cm}P{1.3cm}P{1.3cm}P{1.3cm}P{1.6cm}P{2.4cm}}
\hline
Source ID & RA & Dec & \textit{J} & $\beta$ & [3.6]$-$[4.5] & ib945$-$\textit{Y} & Exp. Time [s] & Avg. Seeing [arcsec] \Tstrut\Bstrut \\[2pt]
\hline
COS-788571 & 09:59:21.68 & +02:14:53.02 & 25.26$^{+0.11}_{-0.10}$ & $-$2.11$\pm$0.55 & $-$0.92$^{+0.21}_{-0.24}$ & 0.17$^{+0.23}_{-0.21}$ & 18900 & 0.92 \Tstrut{} \\[4pt]
COS-851423 & 09:59:11.46 & +02:18:10.42 & 25.90$^{+0.22}_{-0.18}$ & $-$2.64$\pm$0.39 & $-$0.72$^{+0.36}_{-0.48}$ & 0.19$^{+0.27}_{-0.25}$ & 91900 & 0.97 \\[4pt]
COS-854905 & 09:59:09.13 & +02:18:22.38 & 25.75$^{+0.27}_{-0.22}$ & $-$1.95$\pm$0.24 & $-$0.44$^{+0.31}_{-0.36}$ & 0.20$^{+0.34}_{-0.31}$ & 91900 & 0.97 \\[4pt]
COS-856875 & 09:58:45.34 & +02:18:28.87 & 25.63$^{+0.29}_{-0.23}$ & $-$2.06$\pm$0.39 & $-$0.57$^{+0.38}_{-0.44}$ & $-$0.96$^{+0.26}_{-0.32}$ & 73100 & 0.98 \\[4pt]
COS-857605 & 09:59:12.35 & +02:18:28.86 & 25.76$^{+0.23}_{-0.19}$ & $-$1.38$\pm$0.27 & $-$0.45$^{+0.27}_{-0.34}$ & 0.67$^{+0.42}_{-0.35}$ & 65900 & 0.97 \\[4pt]
COS-940214 & 09:59:06.73 & +02:22:45.93 & 26.27$^{+0.45}_{-0.31}$ & $-$2.77$\pm$0.54 & $<-1.26$ & $-$0.87$^{+0.36}_{-0.44}$ & 18900 & 0.92 \\[4pt]   
COS-955126 & 09:59:23.62 & +02:23:32.73 & 25.38$^{+0.24}_{-0.20}$ & $-$2.44$\pm$0.13 & $-$0.94$^{+0.33}_{-0.44}$ & $-$0.16$^{+0.24}_{-0.26}$ & 84700 & 0.95 \\[4pt]
COS-957862 & 09:59:17.62 & +02:23:41.24 & 25.62$^{+0.17}_{-0.15}$ & $-$2.12$\pm$0.30 & - & 1.38$^{+1.24}_{-0.59}$ & - & - \\[4pt]       
COS-1009842 & 09:59:06.33 & +02:26:30.48 & 26.22$^{+0.26}_{-0.21}$ & $-$2.59$\pm$0.44 & $-$0.61$^{+0.42}_{-0.53}$ & $-$0.74$^{+0.28}_{-0.31}$ & 91900 & 0.97 \\[4pt]
COS-1038989 & 09:59:01.40 & +02:28:02.28 & 25.92$^{+0.23}_{-0.19}$ & $-$1.43$\pm$0.30 & $-$0.97$^{+0.56}_{-1.07}$ & 0.27$^{+0.53}_{-0.43}$ & - & - \\[4pt]                                           
COS-1048848 & 09:59:09.76 & +02:28:32.95 & 26.09$^{+0.27}_{-0.22}$ & $-$2.43$\pm$0.34 & $<-0.13$ & 0.59$^{+0.55}_{-0.41}$ & 91900 & 0.97 \\[4pt]
COS-1053257 & 09:58:46.20 & +02:28:45.76 & 24.79$^{+0.08}_{-0.07}$ & $-$2.02$\pm$0.20 & $-$0.33$^{+0.33}_{-0.36}$ & 0.70$^{+0.23}_{-0.20}$ & 91900 & 0.97 \\[4pt]  
\hline
\end{tabular}
\end{table*}

\begin{figure}
\includegraphics{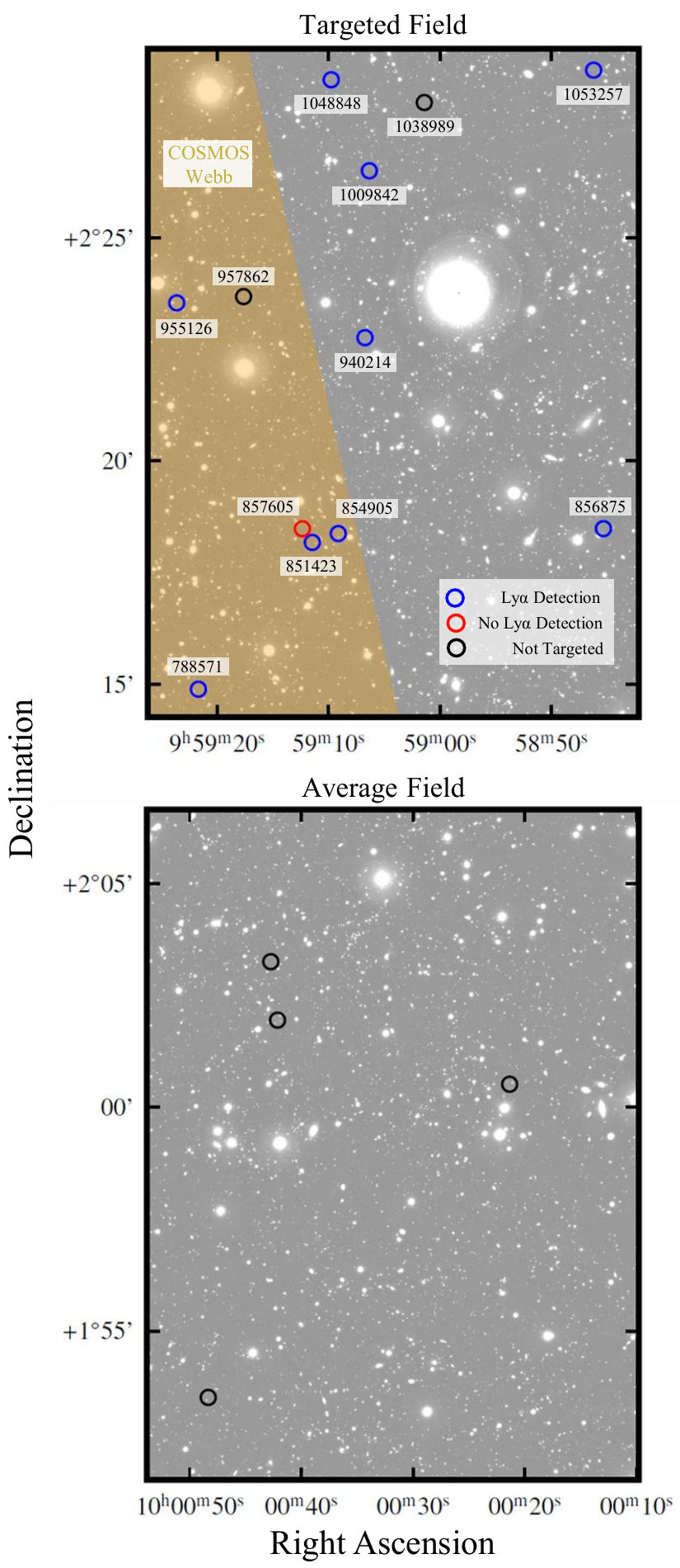}
\caption{The top panel shows the 11$\times$15 arcmin$^2$ area of COSMOS considered in this work. We mark the positions of $z\simeq6.6-6.9$ Lyman-break sources with an MMT/Binospec \Lya{} detection (see Fig. \ref{fig:LyA_Detections}) with blue circles. Those targeted with Binospec and not detected in \Lya{} are shown with red circles, while the remaining sources not targeted with Binospec are shown with black circles. The UltraVISTA \textit{J}-band image is shown in the background with the footprint of the Cycle 1 \JWST{} COSMOS-Webb survey \citep{COSMOSWebbProposal} shown in orange. The bottom panel shows a sub-field of COSMOS with the same area, but with the average surface density of $z\simeq6.6-6.9$ sources we identify across COSMOS. The surface density in our targeted field is 2.85$\times$ this average value, suggesting the presence of a strong, large-scale ($\sim$5 physical Mpc) galaxy overdensity.}
\label{fig:overdensity}
\end{figure}

\section{Sample and Observations} \label{sec:observations}

We are conducting a wide-area (7 deg$^2$) spectroscopic survey targeting \Lya{} emission in Lyman-break selected galaxies at $z\sim7$.
One of the goals of this survey is to identify large ionized bubbles and quantify their local galaxy overdensities.
In \citetalias{Endsley2021_LyA}, we described the first results from this campaign where we detected \Lya{} emission from three UV-bright systems in COSMOS with very similar redshifts ($\zLya{} = 6.75-6.81$) and relatively small projected separations ($<$5 arcmin).
We suggested that these systems may trace a large ($R\gtrsim2$ physical Mpc) ionized bubble formed by a local overdensity, as potentially evidenced by the high surface density of $z\simeq6.8$ galaxy candidates in the surrounding region. 
To further characterize this field, we have since conducted ultra-deep (up to 25 hour) \Lya{} spectroscopy of additional $z\simeq7$ candidates in the surrounding 11$\times$15 arcmin$^2$ area.
With this ultra-deep data, we are able to measure \Lya{} fluxes and redshifts for more sources and thus better quantify both the local \Lya{} EW distribution and galaxy overdensity.
Below, we describe our target selection and the spectroscopic observations.

The wide-area (1.5 deg$^2$) COSMOS field has been imaged with a large number of optical and near-infrared filters, enabling robust identification of the Lyman-alpha break at high redshifts.
The Hyper Surpime-Cam (HSC) Subaru Strategic Program (HSCSSP; \citealt{Aihara2018,Aihara2019}) provides optical imaging in the \textit{g}, \textit{r}, \textit{i}, \textit{z}, and \textit{y} broad-band filters as well as nb816 and nb921 narrow-band filters.
Additionally, COSMOS has been imaged in the HSC nb718, ib945, and nb973 filters with the Cosmic HydrOgen Reionization Unveiled with Subaru (CHORUS) survey \citep{Inoue2020}.
In the near-infrared, the UltraVISTA survey \citep{McCracken2012} has imaged COSMOS with the \textit{Y}, \textit{J}, \textit{H}, and \textit{K}$_s$ broad-band filters of VISTA/VIRCam.
We utilize PDR2 data of HSCSSP, PDR1 data of CHORUS, and DR4 data of UltraVISTA which are all calibrated to the \textit{Gaia} astrometric reference frame.

Our selection of $z\simeq7$ galaxies follows that described in \citet[][hereafter \citetalias{Endsley2021_OIII}]{Endsley2021_OIII} and \citetalias{Endsley2021_LyA}.
Briefly, we first extract sources from a $\chi^2$ \textit{yYJHK}$_s$ detection image \citep{Szalay1999} and apply the color cuts \textit{z}$-$\textit{y}$>$1.5, \textit{z}$-$\textit{Y}$>$1.5, nb921$-$\textit{Y}$>$1.0, and \textit{y}$-$\textit{Y}$<$0.4.
The enforcement of strong dropouts in \textit{z} and nb921 identifies Lyman-alpha breaks at $z\gtrsim6.6$ while the relatively flat \textit{y}$-$\textit{Y} color limits the selection window to $z\lesssim6.9$.
To minimize low-redshift contaminants, we also apply the cuts S/N(\textit{g})$<$2 and S/N(\textit{r})$<$2.
Galactic brown dwarfs are removed from the sample by enforcing \textit{Y}$-$\textit{J}$<$0.45 or (\textit{J}$-$\textit{H}$>$0 and \textit{J}$-$\textit{K}$_s\!\!>$0).
To ensure each galaxy candidate is real, we require S/N$>$3 detections in \textit{y}, \textit{Y}, and \textit{Y}, as well as a S/N$>$5 detection at in least one of those three bands.
All HSC and VIRCam photometry are calculated in 1.2 arcsec diameter apertures with aperture corrections determined from the curve of growth of nearby stars.

We identify a total of twelve $z\simeq6.6-6.9$ Lyman-break galaxies in an 11$\times$15 arcmin$^2$ region of COSMOS surrounding the tentative ionized bubble reported in \citetalias{Endsley2021_LyA}.
This 165 arcmin$^2$ area is fully contained within one of the ultra-deep UltraVISTA stripes, enabling us to select $z\simeq6.6-6.9$ galaxies as faint as $J \sim 26$.
The surface density of $z\simeq6.6-6.9$ candidates in this region (0.0727 arcmin$^{-2}$) is 2.85$\times$ the average across the full 0.73 deg$^2$ area covered by the ultra-deep UltraVISTA stripes (0.0255 arcmin$^{-2}$; see Fig. \ref{fig:overdensity}), perhaps indicating a large-scale ($\sim$5 physical Mpc) overdensity.
We list the coordinates and photometric properties of all twelve systems in Table \ref{tab:table1}, where the three $z=6.75-6.81$ \Lya{} emitters reported in \citetalias{Endsley2021_LyA} are COS-940214, COS-955126, and COS-1009842.
All twelve galaxies have \textit{J}-band magnitudes ranging from $m=24.8$ to 26.3, corresponding to absolute UV magnitudes of $-22 \lesssim \Muv{} \lesssim -20.5$ at $z\simeq6.8$.
We measure the rest-UV slopes by fitting $f_{\lambda} \propto \lambda^{\beta}$ to the VIRCam $YJHK_s$ photometry, finding that our galaxies exhibit slopes spanning $-2.8 \leq \beta \leq -1.4$ consistent with values reported for larger samples of similarly bright $z\sim7$ galaxies (e.g. \citealt{Finkelstein2012,Bouwens2014_beta,Bowler2017}; \citealt{Endsley2021_OIII}).
As described above, we have assumed that apertures of 1.2 arcsec diameter (with point-source corrections) capture the large majority of flux coming from our $z\sim7$ galaxies in the HSC and VIRCam imaging.
This assumption is motivated by the expectation that $-22 \lesssim \Muv{} \lesssim -20.5$ galaxies at $z\sim7$ will have half-light radii of $\approx$0.5--1 kpc (i.e. $\approx$0.1--0.2 arcsec; e.g. \citealt{Shibuya2015,CurtisLake2016,Bowler2017}), which are considerably smaller than our adopted 0.6 arcsec aperture radius.
Indeed, if we adopt a larger aperture radius of 0.9 arcsec \citep[e.g.][]{Bowler2014}, the derived near-IR flux densities of our galaxies typically increase by only 10\% (i.e. 0.1 mag).

For the purpose of inferring stellar masses and ionizing properties, we also calculate the \Spitzer{}/IRAC 3.6 and 4.5$\mu$m photometry of our twelve $z\simeq6.6-6.9$ systems.
We use IRAC data from the SPLASH and SMUVS surveys \citep{Steinhardt2014,Ashby2018} and co-add individual exposures using the \textsc{mopex} pipeline \citep{Makovoz2005_MOPEX} as described in \citetalias{Endsley2021_OIII}.
Due to the broad point spread function of IRAC (FWHM$\approx$2 arcsec), we deconfuse the co-added images using morphological priors from the \HST{}/F814W data available over COSMOS \citep{Koekemoer2007,Massey2010} before calculating photometry in 2.8 arcsec diameter apertures (see \citetalias{Endsley2021_LyA} for further details).
Upon visual inspection, we conclude the deconfusion residuals are acceptably smooth (see \citetalias{Endsley2021_LyA}) in all but one of the twelve identified $z\simeq6.6-6.9$ systems in our targeted field.
This galaxy, COS-957862, has a bright neighboring source $\approx$1.5 arcsec to the East (i.e. within the IRAC FWHM) whose mid-infrared flux profile is not precisely matched by the F814W prior, resulting in a systematic pixel S/N offset within the aperture.
Accordingly, we do not use the IRAC photometry for COS-957862.
The [3.6]$-$[4.5] colors of the eleven remaining sources are all blue with several displaying [3.6]$-$[4.5]$\lesssim$1 (see Table \ref{tab:table1}), suggesting generally strong \OIIIHb{} emission as expected among $z\sim7$ galaxies (\citealt{Smit2014,Smit2015}; \citetalias{Endsley2021_OIII}). 
We come back to discuss the implied physical properties of these systems in \S\ref{sec:BEAGLE}.

We have thus far obtained \Lya{} spectra for ten of the twelve $z\simeq6.8$ galaxies described above (all but COS-957862 and COS-1038989).
Spectroscopic observations were conducted using MMT/Binospec which is a wide-area (240 arcmin$^2$ field of view) multi-object optical spectrograph \citep{Fabricant2019}.
We used the 600 l/mm grating for all observations, enabling sensitive spectroscopic coverage from $\approx$0.7--1.0$\mu$m with moderately-high resolution ($R\approx4400$).
Due to variable seeing and sky transparency between our Binospec exposures, we reduce each frame separately using the publicly available pipeline \citep{Kansky2019} and then co-add the exposures following the weighting approach of \citet{Kriek2015}.
As described in \citetalias{Endsley2021_LyA}, we obtain 1D spectra using optimal extraction \citep{Horne1986} and determine the absolute flux calibration from the spectra of stars on each mask.
We apply slit loss corrections using the size-luminosity relation of \citet{CurtisLake2012} which results in modest correction factors of 5--8\% for all galaxies in our sample given their expected small size (0.1--0.2 arcsec radii).
These slit loss correction factors change very little ($<$1\%) if we instead compute photometry in larger 1.8 arcsec diameter apertures.
The total exposure time for each of the ten targeted $z\simeq6.6-6.9$ galaxies ranges from 18900--91900 seconds (5.2--25.5 hours) with co-added average seeing between 0.9--1.0 arcsec (see Table \ref{tab:table1}).

\begin{table*}
\centering
\caption{Information on confident ($>$7$\sigma$) \Lya{} detections in our targeted $z\simeq6.6-6.9$ galaxy sample. The line properties (redshift, FWHM, flux, and EW) are determined by fitting a truncated Gaussian (Eq. \ref{eq:trunc_gauss}) to each observed 1D profile after masking skyline-contaminated regions of the spectra (see Fig. \ref{fig:LyA_Detections}). In the final two columns, we report the EWs and fluxes derived by combining our measured \Lya{} redshifts with the ib945$-$\textit{Y} color of each source (see Eq. \ref{eq:EWphot}). No systematic offset is apparent between the two methods. We adopt the fluxes and EWs from the line-fitting approach as fiducial.}
\begin{tabular}{P{1.8cm}P{1.5cm}P{0.8cm}P{1.5cm}P{2.1cm}P{1.5cm}P{1.5cm}P{2.1cm}} 
\hline
Source ID & \zLya{} & S/N & FWHM & Flux & EW & EW$_{\mathrm{ib}945-Y}$ & Flux$_{\mathrm{ib}945-Y}$ \Tstrut{} \\
 & & & [km s$^{-1}$] & [10$^{-18}$ erg/s/cm$^2$] & [\AA{}] & [\AA{}] & [10$^{-18}$ erg/s/cm$^2$] \Bstrut{}\\
\hline 
COS-788571 & $6.882^{+0.001}_{-0.000}$ & 16.5 & $272^{+31}_{-29}$ & $15.1^{+1.2}_{-1.2}$ & $28.6^{+4.3}_{-3.6}$ & $34.4\pm8.4$ & $18.2\pm4.9$ \Tstrut{} \\[4pt]
COS-851423 & $6.701^{+0.001}_{-0.001}$ & 11.0 & $260^{+102}_{-57}$ & $5.2^{+1.1}_{-0.7}$ & $11.0^{+3.0}_{-2.1}$ & $2.6\pm9.9$ & $1.2\pm4.8$ \\[4pt]
COS-854905 & $6.702^{+0.002}_{-0.002}$ & 9.4 & $542^{+197}_{-170}$ & $4.8^{+1.4}_{-1.2}$ & $8.8^{+3.3}_{-2.4}$ & $2.1\pm12.0$ & $1.2\pm6.7$ \\[4pt]
COS-856875 & $6.732^{+0.002}_{-0.002}$ & 15.2 & $331^{+54}_{-49}$ & $18.5^{+5.1}_{-4.7}$ & $49.0^{+22.0}_{-15.0}$ & $67.4\pm24.6$ & $25.6\pm11.5$ 
\\[4pt]
COS-940214 & $6.748^{+0.001}_{-0.002}$ & 9.4 & $211^{+33}_{-27}$ & $17.3^{+3.9}_{-2.9}$ & $65.3^{+33.5}_{-18.6}$ & $64.4\pm31.6$ & $17.5\pm10.5$ \\[4pt]
COS-955126 & $6.814^{+0.000}_{-0.000}$ & 21.0 & $403^{+32}_{-32}$ & $13.0^{+0.9}_{-0.9}$ & $19.5^{+4.5}_{-3.2}$ & $38.8\pm13.2$ & $25.8\pm10.1$ \\[4pt]
COS-1009842 & $6.759^{+0.000}_{-0.000}$ & 24.9 & $258^{+23}_{-21}$ & $12.8^{+0.9}_{-0.8}$ & $43.7^{+12.8}_{-8.3}$ & $64.1\pm23.9$ & $18.8\pm8.3$ \\[4pt]
COS-1048848 & $6.847^{+0.001}_{-0.001}$ & 15.1 & $297^{+80}_{-54}$ & $6.6^{+1.1}_{-0.8}$ & $20.6^{+5.8}_{-4.1}$ & $13.4\pm10.3$ & $4.3\pm3.5$ \\[4pt]
COS-1053257 & $6.750^{+0.002}_{-0.002}$ & 7.5 & $149^{+56}_{-51}$ & $3.7^{+3.3}_{-1.8}$ & $3.6^{+3.3}_{-1.8}$ & $-3.4\pm4.2$ & $-3.5\pm4.2$ \\[4pt]
\hline
\end{tabular}
\label{tab:LyA}
\end{table*}

\begin{figure*}
\includegraphics{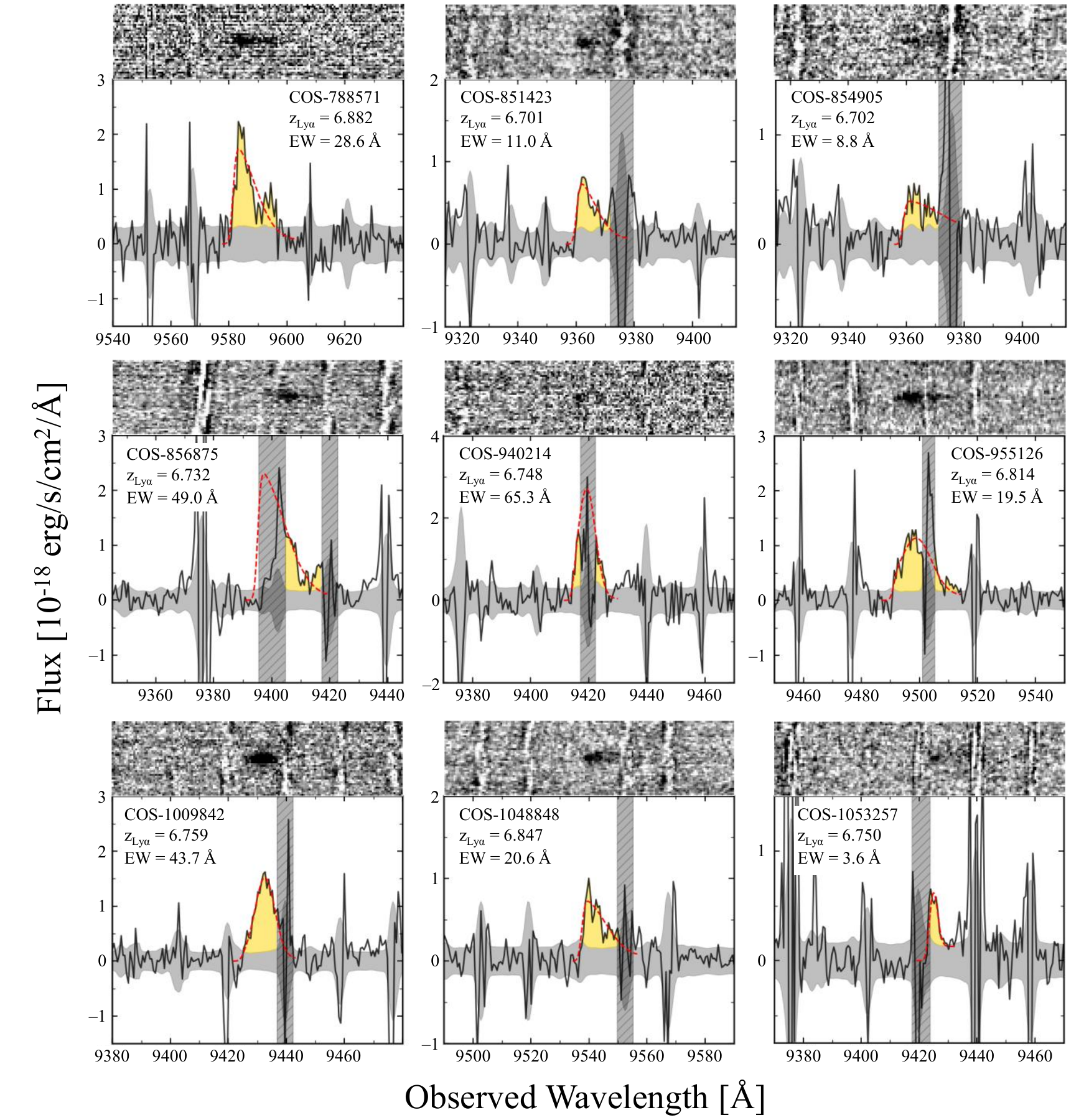}
\caption{Lyman-alpha detections of nine of twelve $z\simeq6.6-6.9$ Lyman-break galaxies identified across an 11$\times$15 arcmin$^2$ region of COSMOS. 2D signal-to-noise maps are shown in the top sub-panels (black is positive) while the 1D extracted spectra are shown on the bottom with the 1$\sigma$ error in grey. All detections have S/N=7.5--24.9 where these values are calculated by integrating the 1D profiles after masking regions contaminated by skylines (shaded hatched regions in lower panels). The best-fitting truncated Gaussian profiles (see Eq. \ref{eq:trunc_gauss}) are shown with dashed red lines and were fit excluding skyline-impacted regions.} 
\label{fig:LyA_Detections}
\end{figure*}

\section{Results} \label{sec:results}

In this section, we first present results from our MMT/Binospec \Lya{} spectra of UV-bright $z\simeq6.6-6.9$ galaxies surrounding the tentative ionized bubble identified in \citetalias{Endsley2021_LyA} (\S\ref{sec:spectra}).
We then infer the physical properties of each galaxy by fitting their photometry with a photoionization model (\S\ref{sec:BEAGLE}), helping inform how efficiently these sources produce and emit \Lya{} photons.

\subsection{MMT/Binospec Spectra} \label{sec:spectra}

We have targeted \Lya{} emission from ten Lyman-break $z\simeq6.6-6.9$ galaxies spanning an 11$\times$15 arcmin$^2$ area around the potential ionized bubble reported in \citetalias{Endsley2021_LyA} (Fig. \ref{fig:overdensity}).
Our ultra-deep MMT/Binospec spectra reveal an emission line detection in nine of ten targeted sources (see Fig. \ref{fig:LyA_Detections}).
We interpret these emission lines as \Lya{} for the following reasons. 
First, the observed wavelength of each detected feature is consistent with a \Lya{} solution given the expected redshift range of our Lyman-break selection criteria ($z\simeq6.6-6.9$).
Second, none of the emission features appear to be an [OII]$\lambda$3727,3729 doublet (i.e. two symmetric emission lines with an observed separation of $\approx$6.8 \AA{}), and we moreover find that [OII] solutions are inconsistent with the photometric data of each source using the SED fitting procedure described in \S\ref{sec:BEAGLE}.
We consider [OII] as the most likely alternative solution to \Lya{} since, if at low redshift, a Balmer break would likely be necessary to produce the very strong \textit{z} and nb921 dropouts of each galaxy.
Finally, we note that no other emission features are detected in these spectra, indicating that H$\beta$ and [OIII]$\lambda$4959,5007 solutions are unlikely.
The \Lya{} detections of four galaxies (COS-788571, COS-940214, COS-955126, and COS-1009842) were reported in \citetalias{Endsley2021_LyA}, though we have since obtained much deeper spectra (i.e. 3--5$\times$ longer integration) on the latter two galaxies.

The significance of each \Lya{} detection ranges from 7.5--24.9$\sigma$ (see Table \ref{tab:LyA}).
These values are calculated by integrating the 1D line profile after masking regions contaminated by skylines, as determined from visual inspection to the 2D data (see Fig. \ref{fig:LyA_Detections}).
Because skylines appear to be obscuring significant portions of \Lya{} flux for some sources (e.g. COS-856875 and COS-940214), we determine the line properties (e.g. redshift and flux) of all galaxies by fitting their 1D spectra after masking the skyline-contaminated regions.
We fit each spectra to a truncated Gaussian profile that accounts for varying degrees of line asymmetry seen in our sample (for example, compare the line shapes of COS-1009842 and COS-1048848 in Fig. \ref{fig:LyA_Detections}):
\begin{equation} \label{eq:trunc_gauss}
F(\lambda) = 
\begin{cases} 
      0 & \mathrm{for}\ \lambda < \lambda_{\mathrm{cut}} \\
      A \ e^{-\left(\lambda-\lambda_0\right)^2/2\sigma^2} + C & \mathrm{for}\ \lambda 
      \geq \lambda_{\mathrm{cut}}
\end{cases}.
\end{equation}
Here $A$, $\lambda_0$, and $\sigma$ are the typical definitions of amplitude, peak wavelength, and standard deviation of a Gaussian profile, while $C$ is the continuum flux density set equal to the value from the VIRCam \textit{Y}-band photometry. 
The value $\lambda_{\mathrm{cut}}$ defines the wavelength below which the \Lya{} optical depth rises sharply (due to either CGM or IGM absorption), and is parametrized as $\lambda_{\mathrm{cut}} \equiv \lambda_0 - k\sigma$ where we allow $0\leq k \leq3$ when fitting the data.
A value of $k=0$ yields a line with strong asymmetry while $k=3$ results in an almost perfectly symmetric profile.
During each fit, we convolve the profile from Eq. \ref{eq:trunc_gauss} with the instrumental response ($R=4400$) and use \textsc{emcee} \citep{ForemanMackey2013_emcee} to determine confidence intervals on the four free parameters: $A$, $\lambda_0$, $\sigma$, and $k$.
The best-fitting \Lya{} profiles are shown in Fig. \ref{fig:LyA_Detections}.

The peak wavelengths of the nine \Lya{} emission lines correspond to redshifts between $\zLya{} = 6.701-6.882$ (see Table \ref{tab:LyA}) as expected given our $z\simeq6.6-6.9$ Lyman-break selection.
After correcting for instrumental broadening, we find that the FWHM of each \Lya{} profile ranges between 149 and 542 km s$^{-1}$, with a median value of 272 km s$^{-1}$ (Table \ref{tab:LyA}).
This is consistent with the median FWHM of $\approx$270 km s$^{-1}$ measured from individual $z=7-8$ \Lya{} spectra in \citet{Jung2020}, and is slightly broader than the FWHM$\approx$220 km s$^{-1}$ obtained from the stacked spectra of $z\sim7$ \Lya{} emitters in \citet{Pentericci2018}.
The line fluxes of our Binospec-detected sample span (3.7--17.3)$\times$10$^{-18}$ erg/s/cm$^2$ (Table \ref{tab:LyA}) where we have subtracted off the continuum with flux density assumed equal to that from the \textit{Y}-band photometry.
The rest-frame EWs derived from these line fluxes span 3.6--65.3 \AA{} (see Table \ref{tab:LyA}), where four galaxies are found to exhibit strong \Lya{} emission (EW$>$25 \AA{}).

To check that the \Lya{} fluxes and EWs derived from our line-fitting procedure are reasonable, we also derive the EWs using an alternative approach that combines both the spectra and photometry.
In much the same way that $z\sim7-8$ \OIIIHb{} EWs can be inferred from IRAC colors, \Lya{} EWs can also be estimated from optical/near-infrared photometry if the redshift is known.
At the redshifts of our galaxies ($z=6.70-6.88$), the \Lya{} line falls within the HSC ib945 band and thus a blue ib945$-$\textit{Y} color will be seen in sources with sufficiently high \Lya{} EW.
As a competing effect, the \Lya{} break also shifts into ib945 at $z>6.65$ and thus will redden the ib945$-$\textit{Y} color to a larger degree at higher redshifts. 
Since we know the redshift from our Binospec spectra, we can calculate the rest-frame \Lya{} EW as
\begin{equation}
\label{eq:EWphot}
\mathrm{EW}_{\mathrm{ib}945-Y} = \frac{\mathrm{BW_{ib945}} \times 10^{-0.4\left(\mathrm{ib945}-Y\right)} - \int_{\lambda_{\Lya{}}}^{\lambda_{\mathrm{max}}} T_{\mathrm{ib945}}(\lambda) d\lambda}{\left(1+\zLya{}\right) T_{\mathrm{ib945}}(\lambda_{\Lya{}})} .
\end{equation}
In this equation, $\mathrm{BW_{ib945}}$ is the bandwidth of the ib945 filter in Angstroms, $T_{\mathrm{ib945}}(\lambda)$ is the filter throughput as a function of observed wavelength (normalized to a maximum value of unity), and $\lambda_{\mathrm{max}}$ is the maximum wavelength covered by ib945.
The \Lya{} EWs (and corresponding fluxes) derived using this secondary approach are listed in the last two columns of Table \ref{tab:LyA}.
These values all agree with those derived using the line-fitting procedure within uncertainties and no systematic offset is apparent between the two methods.
We adopt the fluxes and EWs from the line-fitting approach as fiducial because this method most directly utilizes the \Lya{} spectra and results in more precise EW measurements for all sources.

We observe a range of \Lya{} profiles among our $z\simeq6.8$ galaxies.
Most sources show asymmetric \Lya{} lines (e.g. COS-788571, COS-851423, COS-854905, COS-955126, and COS-1048848; see Fig. \ref{fig:LyA_Detections}), as are commonly seen from galaxies due to resonant nature of \Lya{} emission \citep[e.g.][]{Kunth1998,Hu2004,Steidel2010}.
However, one of our galaxies (COS-1009842) appears to show a highly symmetric \Lya{} velocity profile with $k>2$ at 92\% confidence (Fig. \ref{fig:LyA_Detections}).
Data of COS-1009842 indicate that this symmetric emission feature is unlikely to be a strong non-resonant optical line (i.e. H$\beta$, H$\alpha$, [OIII]$\lambda$4959, or [OIII]$\lambda$5007) at lower redshift.
Using the SED fitting procedure described in \S\ref{sec:BEAGLE}, we find that the photometry of COS-1009842 strongly disfavors such low-redshift solutions (reduced $\chi^2 > 2.2$) while the \Lya{} solution well matches the data (reduced $\chi^2 = 0.6$).
Furthermore, if this feature were H$\beta$, [OIII]$\lambda$4959, or [OIII]$\lambda$5007, at least one of the other two lines would lie in a skyline-free region of the spectrum where no emission is detected.
We discuss what the symmetric \Lya{} velocity profile of COS-1009842 may imply for its ionizing efficiency and surrounding IGM in \S\ref{sec:analysis}.

\begin{table*}
\centering
\caption{Inferred properties of the twelve $z\simeq6.6-6.9$ UV-bright galaxies identified within our targeted region of COSMOS. These properties were obtained using the \textsc{beagle} SED fitting tool which adopts photoionization models of star-forming galaxies. The fiducial values and errors are determined by calculating the median and inner 68\% confidence interval values marginalized over the posterior probability distribution function output by \textsc{beagle}. We report spectroscopic redshifts for sources with a \Lya{} detection and photometric redshifts for those without. In the final column, we report the estimated radius of the HII region generated by each individual source (see Eq. \ref{eq:bubbleSize}). We divide the table into those targeted with MMT/Binospec (top) and those not targeted (bottom). Due to the strong IRAC confusion for COS-957862 (see \S\ref{sec:observations}), we do not report inferred properties strongly dependent on rest-optical data for this source.}
\begin{tabular}{P{1.8cm}P{1.2cm}P{1.1cm}P{1.5cm}P{1.3cm}P{1.5cm}P{1.8cm}P{1.5cm}P{1.2cm}} 
\hline
Source ID & Redshift & \Muv{} & $\tau_{_V}$ & \logten{} \Mstar{} & sSFR & [OIII]+H$\beta$ EW & \logten{} \xiion{} & $R_{_{\mathrm{HII}}}$ \Tstrut{} \\
 &  &  & & [\Msol{}] & [Gyr$^{-1}$] & [\AA{}] & [erg$^{-1}$ Hz] & [pMpc] \Bstrut{} \\
\hline
\multicolumn{9}{c}{Targeted with Binospec} \Tstrut{} \\[4pt]
COS-788571 & 6.882 & $-21.5^{+0.1}_{-0.1}$ & $0.01^{+0.04}_{-0.01}$ & $9.0^{+0.4}_{-0.3}$ & $44^{+49}_{-31}$ & $4040^{+1930}_{-1730}$ & $26.32^{+0.18}_{-0.18}$ & 0.95 \\[4pt]
COS-851423 & 6.701 & $-21.1^{+0.1}_{-0.1}$ & $0.01^{+0.04}_{-0.01}$ & $8.3^{+0.9}_{-0.3}$ & $85^{+107}_{-78}$ & $1020^{+770}_{-500}$ & $25.72^{+0.14}_{-0.14}$ & 0.85 \\[4pt]
COS-854905 & 6.702 & $-21.3^{+0.1}_{-0.1}$ & $0.07^{+0.11}_{-0.06}$ & $8.8^{+0.9}_{-0.5}$ & $67^{+110}_{-62}$ & $670^{+530}_{-360}$ & $25.85^{+0.21}_{-0.19}$ & 0.90 \\[4pt]
COS-856875 & 6.732 & $-21.0^{+0.2}_{-0.1}$ & $0.02^{+0.06}_{-0.02}$ & $8.5^{+0.7}_{-0.5}$ & $58^{+91}_{-51}$ & $680^{+440}_{-340}$ & $25.68^{+0.16}_{-0.19}$ & 0.84 \\[4pt]
COS-857605 & $6.73^{+0.05}_{-0.04}$ & $-21.3^{+0.1}_{-0.1}$ & $0.22^{+0.07}_{-0.08}$ & $9.1^{+0.6}_{-0.5}$ & $30^{+69}_{-25}$ & $470^{+370}_{-220}$ & $25.95^{+0.16}_{-0.23}$ & 0.91 \\[4pt]
COS-940214 & 6.748 & $-20.4^{+0.2}_{-0.2}$ & $0.01^{+0.04}_{-0.01}$ & $8.0^{+0.4}_{-0.3}$ & $90^{+368}_{-62}$ & $3170^{+1980}_{-2010}$ & $25.88^{+0.16}_{-0.18}$ & 0.69 \\[4pt]
COS-955126 & 6.814 & $-21.5^{+0.1}_{-0.1}$ & $0.01^{+0.05}_{-0.01}$ & $8.5^{+0.8}_{-0.3}$ & $99^{+120}_{-90}$ & $1500^{+930}_{-630}$ & $25.85^{+0.16}_{-0.12}$ & 0.96 \\[4pt]
COS-1009842 & 6.759 & $-20.6^{+0.1}_{-0.1}$ & $0.02^{+0.07}_{-0.02}$ & $8.3^{+0.7}_{-0.4}$ & $69^{+111}_{-60}$ & $900^{+720}_{-460}$ & $25.76^{+0.19}_{-0.16}$ & 0.73 \\[4pt]
COS-1048848 & 6.847 & $-20.7^{+0.2}_{-0.1}$ & $0.01^{+0.02}_{-0.01}$ & $8.3^{+0.4}_{-0.4}$ & $45^{+65}_{-33}$ & $410^{+520}_{-230}$ & $25.46^{+0.24}_{-0.26}$ & 0.74 \\[4pt]
COS-1053257 & 6.750 & $-22.0^{+0.1}_{-0.1}$ & $0.05^{+0.07}_{-0.04}$ & $8.7^{+1.3}_{-0.4}$ & $98^{+132}_{-95}$ & $510^{+410}_{-260}$ & $25.75^{+0.12}_{-0.20}$ & 1.13 \\[4pt]
\hline
\multicolumn{9}{c}{Not targeted with Binospec} \Tstrut{} \\[4pt]
COS-957862 & $6.81^{+0.06}_{-0.06}$ & $-21.2^{+0.1}_{-0.1}$ & $0.02^{+0.06}_{-0.01}$ & - & - & - & - & 0.88 \\[4pt]
COS-1038989 & $6.67^{+0.07}_{-0.06}$ & $-20.8^{+0.1}_{-0.1}$ & $0.11^{+0.12}_{-0.10}$ & $8.5^{+0.7}_{-0.4}$ & $87^{+126}_{-74}$ & $1230^{+1050}_{-680}$ & $26.07^{+0.18}_{-0.20}$ & 0.80 \\[4pt]
\hline
\end{tabular}
\label{tab:galaxy_properties}
\end{table*}

\subsection{Photoionization Modeling} \label{sec:BEAGLE}

A primary goal of this work is to test if the  $z\simeq6.6-6.9$ galaxies targeted in this paper have enhanced \Lya{} emission, as would be expected if they are situated in a large ionized bubble.
Since the strength of \Lya{} could alternatively be boosted by properties internal to galaxies,  we must first consider how the properties of these systems may impact their \Lya{} output. 
To this end, we infer the physical properties of the $z\simeq6.6-6.9$ galaxies in our targeted field by fitting their photometry with the \textsc{beagle} SED fitting tool (v0.20.4; \citealt{Chevallard2016}). 
We will compare the inferred properties of these systems to the larger (N=22) Binospec-targeted sample of equally luminous ($\Muv{} \leq -20.4$) $z\simeq6.6-6.9$ galaxies described in \citetalias{Endsley2021_LyA}.

\textsc{beagle} utilizes the \citet{Gutkin2016} photoionization models of star-forming galaxies which include both stellar and nebular emission, computed by processing the latest version of the \citet{BruzualCharlot2003} stellar population synthesis models through \textsc{cloudy} \citep{Ferland2013}.
In \textsc{beagle}, the posterior probability distributions of galaxy properties are derived from the input photometry using the Bayesian \textsc{multinest} algorithm \citep{Feroz2008,Feroz2009}. 
Our fitting procedure matches that described in \citetalias{Endsley2021_LyA} with the exception that we fold in the narrow/intermediate-band CHORUS data for sources lacking a spectroscopic detection, thereby providing more precise photometric redshifts (we also update the fits of the larger \citetalias{Endsley2021_LyA} sample with this new photometry).
We employ a \citet{Chabrier2003} IMF with mass limits of 0.1--300 \Msol{}, adopt an SMC dust prescription \citep{Pei1992}, and assume a delayed star formation history (SFR $\propto t e^{-t/\tau}$) with an allowed superimposed recent (1--10 Myr) burst. 
Such bursts may have recently occurred within at least a subset of our UV-bright $z\simeq6.6-6.9$ systems showing very blue IRAC colors ([3.6]$-$[4.5]$\lesssim -1$) which are indicative of high-EW \OIIIHb{} emission (\citetalias{Endsley2021_OIII}).
Because the transmission of \Lya{} photons is not a free parameter within \textsc{beagle}, we remove \Lya{} emission from the model templates and only fit data redward of the Lyman-alpha break for sources with a redshift measurement.

\begin{figure*}
\includegraphics{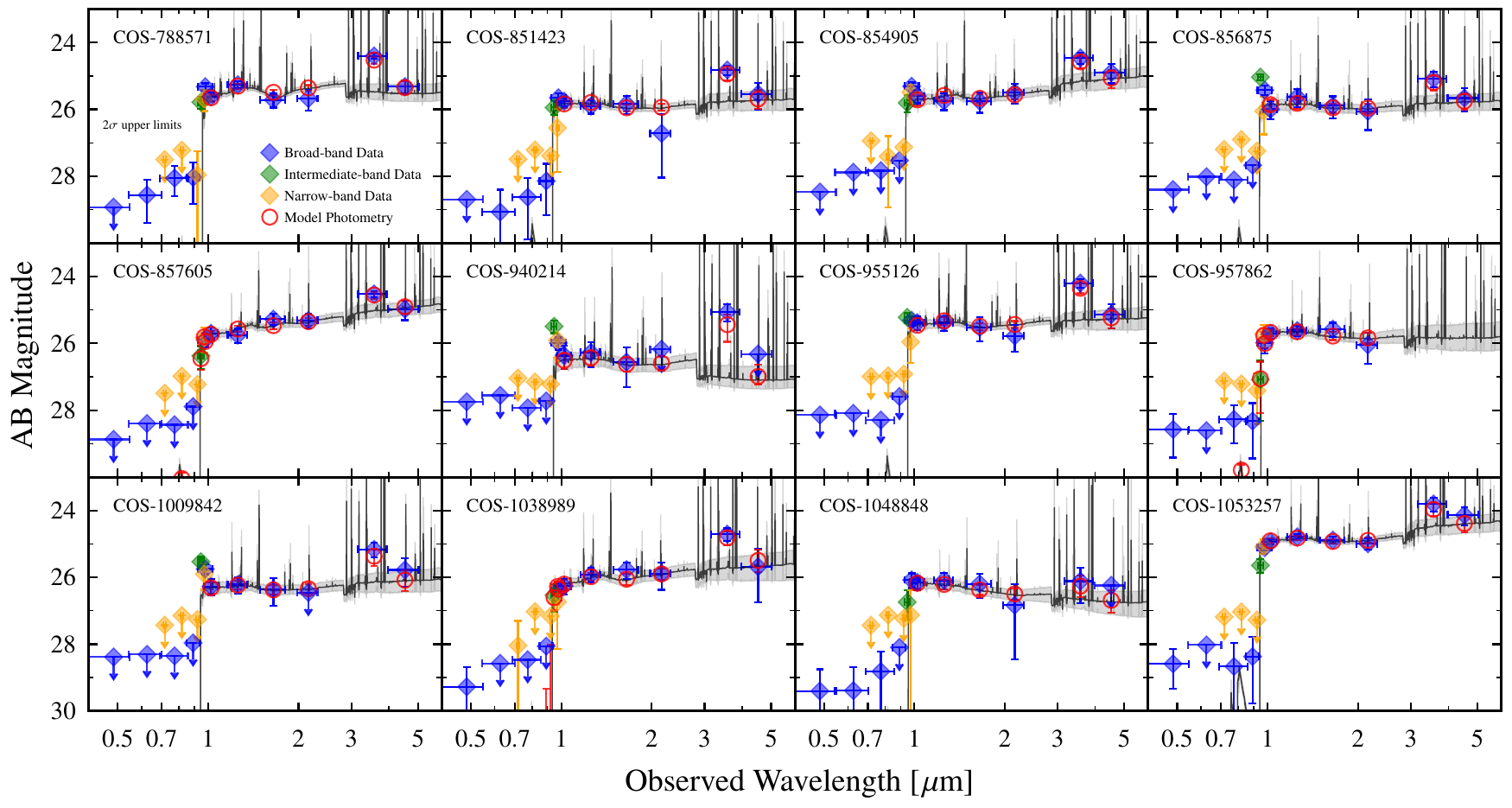}
\caption{Optical through mid-infrared (0.3--5$\mu$m) photometric SEDs of the twelve $z\simeq6.6-6.9$ galaxies in our targeted region of COSMOS. The blue, green, and orange diamonds show the broad-, intermediate-, and narrow-band data, respectively with 2$\sigma$ upper limits shown in cases of non-detections. The fitted median and 68\% confidence interval SEDs from \textsc{beagle} are shown with the black line and gray shaded regions, respectively, in each panel. The associated model photometry is shown with red circles for fitted bands. We only fit data redward of the \Lya{} line for spectroscopically detected sources (see text).}
\label{fig:SEDs}
\end{figure*}

The resulting SED fits and inferred physical properties of each $z\simeq6.6-6.9$ galaxy targeted in this paper are shown in Fig. \ref{fig:SEDs} and Table \ref{tab:galaxy_properties}, respectively. 
For completeness, we also include the two color-selected sources that lie in the same small area as the other sources in this paper but have yet to be targeted with MMT/Binospec.
All twelve galaxies have absolute UV magnitudes spanning $-22.0 \leq \Muv{} \leq -20.4$, corresponding to 0.9--4.0$\times$ the characteristic UV luminosity at $z\sim7$ ($\mathrm{M_{UV}^{\ast}} = -20.5$; \citealt{Harikane2021}).
Among the eleven galaxies with robust IRAC measurements, the inferred stellar masses range from $\logten{}(\Mstar{}/\Msol{})$ = 8.0 to 9.1 with \OIIIHb{} EWs spanning 410--4040 \AA{} (Table \ref{tab:galaxy_properties}) as expected from their blue IRAC colors.
Consistent with these large \OIIIHb{} EWs, we infer high specific star formation rates (sSFRs) ranging from 30 to 99 Gyr$^{-1}$ which are calculated using the average star formation rate over the past 10 Myr.

In the next section, we test for enhanced \Lya{} emission strengths among the ten targeted galaxies that are the subject of this paper. 
To do so, we compare their median \Lya{} EW to that from the larger sample described in \citetalias{Endsley2021_LyA}. 
Our goal in this section is to determine whether we might expect to see substantial variation in the \Lya{} properties of the two samples based solely on differences in their physical properties. 
We first consider the dust content implied by the SED fitting results. 
Dust typically acts to absorb resonantly-scattered \Lya{} photons, and not surprisingly observational measures of dust reddening are well-known to be closely linked to the \Lya{} EW with redder galaxies typically showing weaker \Lya{} emission \citep[e.g.][]{Shapley2003,Pentericci2009,Kornei2010,Stark2010,Schenker2014,Hathi2016,Trainor2016,deBarros2017}.
\textsc{beagle} quantifies the dust content in terms of the V-band dust optical depth parameter, $\tau_{_\mathrm{V}}$, which can be converted into other extinction parameters as, e.g., $E(B-V) = 0.37 \tau_{_\mathrm{V}}$ and $A_{1500} = 5.2 \tau_{_\mathrm{V}}$ using the SMC extinction law adopted in this work.
The SED fits suggest that the median dust content of the ten galaxies targeted in this paper ($\tau_{_\mathrm{V}} = 0.02$) is essentially identical to that of the larger sample from \citetalias{Endsley2021_LyA} ($\tau_{_\mathrm{V}} = 0.03$). 
These values indicate minimal obscuration from dust, suggesting that both samples are likely to have ideal conditions to support the transmission of \Lya{} photons through the gas within the galaxies. 

The production of \Lya{} can also vary considerably within the population. 
This is particularly important for reionization-era galaxies given the range of \OIIIHb{} EWs spanned by the population. 
The ionizing photon production rate (at fixed SFR) varies significantly with \OIIIHb{} EW \citep[e.g.][]{Chevallard2018_z0,Tang2019}, as expected given the close link between the latter quantity and the light-weighted age of the stellar population. 
Assuming case B recombination, variations in ionizing photon production lead to variations in the production of \Lya{}. 
Such dispersion is observed at lower redshift, where the median \Lya{} EW is found to increase with \OIIIHb{} EW \citep[e.g.][]{Du2020,Tang2021_LyA}. 
The median inferred \OIIIHb{} EW of the ten galaxies observed in this work is 790 \AA{} (see Table \ref{tab:galaxy_properties}), only 0.05 dex higher than that of the \citetalias{Endsley2021_LyA} sample (710 \AA{}).
From the lower-redshift studies \citep{Tang2021_LyA}, we expect this will cause a small ($\lesssim$0.05 dex) increase in the \Lya{} EW.

We can also characterize potential variations in the \Lya{} production rate using the ionizing photon production efficiency ($\xiion{}$) implied by the \textsc{beagle} models. 
Here we define $\xiion{}$ as the ratio of the production rate of Lyman-continuum photons ($>$13.6 eV) and the observed UV continuum luminosity at rest-frame 1500 \AA{}.
While other works adopt the ionizing photon production efficiency defined relative to the dust-corrected UV luminosity ($\xi_\mathrm{ion}^{\mathrm{HII}}$) or the intrinsic UV stellar continuum luminosity ($\xi_\mathrm{ion}^{\ast}$, e.g. \citealt{Matthee2017_ionization,Stark2017,Shivaei2018,Chevallard2018_z0,Tang2019}), we here use \xiion{} defined relative to the observed UV luminosity given that the \Lya{} EW is similarly defined.
For reference, our inferred \xiion{} values are typically $\approx$0.05 and $\approx$0.10 dex larger than the inferred $\xi_\mathrm{ion}^{\mathrm{HII}}$ and $\xi_\mathrm{ion}^{\ast}$ values, respectively.
Among the ten galaxies observed in this work, we infer hydrogen ionizing photon production efficiencies of \logxiion{} = 25.46--26.32 with a median \logxiion{} = 25.81 (see Table \ref{tab:galaxy_properties}).
This median \xiion{} is essentially identical to that inferred among the \citetalias{Endsley2021_LyA} galaxies (\logxiion{} = 25.82), suggesting that there are no significant differences in the typical ionizing photon production efficiency between the two samples. 
We note that these median inferred \xiion{} values are moderately ($\approx$0.3 dex) larger than that expected from the $z\sim2$ relation of \citet{Tang2019} connecting ionizing photon production efficiency and [OIII] EW.
Nonetheless, if we instead adopted \xiion{} values using these low-redshift relations, we would still infer very little difference in the typical \Lya{} photon production rates of the two $z\sim7$ samples ($\lesssim$0.03 dex) given that their IRAC colors imply very similar median optical line EWs.
Overall, we do not find any evidence that indicates that the ten galaxies that form the basis of this paper would show significantly different  \Lya{} properties than the larger sample  observed in \citetalias{Endsley2021_LyA}. We will come back to this point in the next section.

\section{Analysis} \label{sec:analysis}

We have two goals in this section. 
First we utilize our Binospec redshift measurements to determine whether our targeted region of COSMOS traces a strong galaxy overdensity, as suggested by the high surface density of $z\simeq6.6-6.9$ Lyman-break systems identified across the corresponding area (\S\ref{sec:overdensity}).
Second, we explore whether our observed galaxies show significantly enhanced \Lya{} emission relative to the average at $z\simeq7$, as would be expected if a very large ionized bubble has formed around these systems (\S\ref{sec:bubble}).

\subsection{A Spectroscopic Overdensity at \boldmath{$z=6.8$}} \label{sec:overdensity}

The high (2.85$\times$ average) surface density of $z\simeq6.6-6.9$ Lyman-break galaxies across our targeted region of COSMOS implies a strong galaxy overdensity (see Fig. \ref{fig:overdensity}).
Equipped with the line-of-sight position information from our \Lya{} detections, we can now spectroscopically constrain this overdensity.
To do so, we compare the number of $\Muv{} \leq -21$ galaxies confirmed to lie in this region to the average number expected from $z\sim7$ UV-luminosity functions derived from wide-area imaging \citep{Bowler2017,Harikane2021}.
While we detect \Lya{} in systems as faint as $\Muv{} = -20.4$ (see Table \ref{tab:galaxy_properties}), we restrict our attention to $\Muv{} \leq -21$ galaxies for this overdensity calculation to avoid luminosity regimes with low completeness. 
Due to the depth of the HSCSSP and UltraVISTA imaging, the completeness of our $z\simeq6.6-6.9$ Lyman-break selection declines sharply at $\Muv{} \gtrsim -21$ ($J \gtrsim 26$) to less than 10\% at $\Muv{} = -20.5$ (see figure 2 of \citetalias{Endsley2021_OIII}).

Of the seven $\Muv{} \leq -21$ Lyman-break galaxies we targeted with Binospec, we have detected \Lya{} from $N = 6$ such systems.
All six of these \Lya{} emitters were selected across an 11$\times$15 arcmin$^2$ area and have redshifts between \zLya{} = 6.701--6.882, indicating that they occupy a volume of $V = 3.5\times4.8\times8.3$ physical Mpc$^3$ = 140 physical Mpc$^3$. 
The average number of $\Muv{} \leq -21$ galaxies expected in this volume is $\langle N \rangle = 1.0^{+0.8}_{-0.5}$ from the \citet{Bowler2017} luminosity function (LF) and $\langle N \rangle = 2.0^{+0.9}_{-0.7}$ from the \citet{Harikane2021} LF.
Because the results from these two LFs are consistent within 1$\sigma$ uncertainties, we adopt the most conservative (i.e. largest) average number from the \citet{Harikane2021} LF as fiducial though include the \citet{Bowler2017} LF values in our reported errors.
Our Binospec data therefore reveal a spectroscopic overdensity of at least $N / \langle N \rangle = 6.0 / 2.0^{\scaleto{+0.9}{4.5pt}}_{\scaleto{-1.5}{4.5pt}} = 3.0^{\scaleto{+9.0}{4.5pt}}_{\scaleto{-0.9}{4.5pt}}$ across the 140 physical Mpc$^3$ volume spanned by our six confirmed $\Muv{} \leq -21$ \Lya{} emitters.
We note that this quoted overdensity is a lower limit since we have not corrected $N$ for either the spectroscopic completeness in detecting \Lya{} nor the photometric completeness in selecting $\Muv{} \leq -21$ $z\simeq6.8$ galaxies within our targeted volume.
The calculated volume overdensity between $z=6.70-6.88$ ($N / \langle N \rangle \gtrsim 3$) is consistent with the 2.85$\times$ average surface density of $z\simeq6.6-6.9$ Lyman-break galaxies identified across our targeted region of COSMOS.

With our \Lya{} detections, we can also measure the physical separations of our galaxies.
Considering now all nine confirmed \Lya{} emitters in our sample, we find that our galaxies are separated by distances between $D = 0.20-8.4$ physical Mpc with a typical separation of 4.3 physical Mpc.
Here, we are ignoring the line-of-sight position uncertainty due to differences in peculiar motions and \Lya{} velocity offsets of our galaxies.
For reference, a combined 200 km s$^{-1}$ difference between two galaxies at $z=6.8$ would yield a derived line-of-sight separation of 0.2 physical Mpc, much smaller than the typical separation of our galaxies.
The two \Lya{} emitters with the largest distance from one another are COS-851423 ($\zLya{} = 6.701$) and COS-788571 ($\zLya{} = 6.882$), where their separation is mainly along the line of sight direction (8.3 physical Mpc).
Accounting for line-of-sight position uncertainties due to peculiar motions ($\sim$0.2 physical Mpc) translates to a small ($\sim$3\%) systematic error in our estimated 140 physical Mpc$^3$ volume above.
The two most closely-separated galaxies in our sample (COS-851423 and COS-854905) have essentially equal \Lya{} redshifts ($z = 6.701-6.702$) and are located only 37 arcsec apart on the sky, translating to a physical separation of $D = 200$ kpc.
Notably, there are two other galaxy pairs in our Binospec-detected sample with fairly small physical separations ($D<2$ Mpc).
COS-940214 and COS-1009842 are located 1.3 physical Mpc apart while COS-1009842 and COS-1053257 are separated by 1.8 physical Mpc.

Given the relatively small physical separation ($D\lesssim1$ Mpc) between some of our galaxies, we consider whether these individual sources could generate ionized bubbles of sufficient size to have overlapped.
We approximate the radius of an HII region, $R_{_{\mathrm{HII}}}$, generated by a single $z\sim7$ galaxy using the equation (e.g. \citealt{Haiman1997})
\begin{equation} \label{eq:bubbleSize}
    \frac{dR_{_{\mathrm{HII}}}}{dt} = \frac{\langle\xiion{}\rangle\ \langle\fesc{}\rangle\ L_{\scaleto{\mathrm{UV}}{4pt}}}{4 \pi R_{_{\mathrm{HII}}}^2\ \bar{n}_{_{\mathrm{HI}}}(z)} + R_{_{\mathrm{HII}}}\ H(z) - R_{_{\mathrm{HII}}}\ \alpha_B\ \bar{n}_{_{\mathrm{HI}}}(z)\ \frac{C_\mathrm{HI}}{3}.
\end{equation}
In this equation, the first and second terms on the right account for bubble growth due to photoionization and the Hubble flow, respectively, while the third term accounts for recombinations in the IGM. 
The terms $\bar{n}_{_{\mathrm{HI}}}(z)$ and $H(z)$ are, respectively, the average HI density and Hubble parameter at the redshift corresponding to the associated time step.
For the case B recombination coefficient, we assume a temperature of $10^4$ K yielding $\alpha_B = 2.59\times10^{-13}$ cm$^3$ s$^{-1}$ \citep{OsterbrockFerland2006} and we set the IGM HI clumping factor, $C_\mathrm{HI}$, to a fixed value of 3 (see e.g. \citealt{Robertson2013}).
Finally, $\langle \xiion{}\rangle$ and $\langle\fesc{}\rangle$ are the average ionizing photon production efficiency and escape fraction, respectively, during the lifetime of the emitting galaxy.

The inferred ages of our galaxies are largely determined by their IRAC colors which probe the strength of \OIIIHb{} emission relative to the underlying optical continuum. A subset of our galaxies show IRAC colors suggesting extremely strong \OIIIHb{} emission (EW$\geq$1500 \AA{}; see Table \ref{tab:galaxy_properties}), thereby implying very young ages ($t\lesssim10$ Myr) assuming a constant star formation history. 
In these systems, it is possible that an older stellar population is being outshined by a recent strong upturn of star formation activity, limiting our knowledge of the true duration of ionizing photon production in these systems (e.g. \citealt{Tang2022}).
We thus adopt a single timescale for star formation (and hence ionizing photon production) for all our galaxies which is estimated from the typical $z\simeq7$ \OIIIHb{} EW of 760 \AA{} (\citetalias{Endsley2021_OIII}).
This EW corresponds to a galaxy age of approximately 200 Myr assuming a constant star formation history with 0.2 solar metallicity, ionization parameter of log $U = -2.5$, and dust-to-metal mass ratio $\xi_d = 0.3$ using the \citet{Gutkin2016} photoionization models of star-forming galaxies.
Over this 200 Myr age, we adopt a time-averaged ionizing photon production efficiency and escape fraction of $\langle \xiion{} \rangle$ = 10$^{25.8}$ erg$^{-1}$ Hz (\citealt{Stark2017,deBarros2019}; \citetalias{Endsley2021_LyA}) and $\langle \fesc{} \rangle$ = 0.2 \citep[e.g.][]{Robertson2013}, respectively.
As discussed in \S\ref{sec:BEAGLE}, low-redshift observations \citep{Tang2019} predict $\approx$2$\times$ lower ionizing photon production efficiencies for our galaxies relative to that inferred from \textsc{beagle}.
However, adopting these lower \xiion{} values would only change our estimated bubble sizes by a modest $\approx$0.1 dex (i.e. $\approx$25\%).
A factor of 2 change in our assumed \fesc{} would have an equally modest effect.

Based on the assumptions above, it is plausible that each of our UV-bright ($\Muv{} \leq -20.4$) $z\simeq6.6-6.9$ galaxies reside in ionized bubbles with radii of $R = 0.69-1.13$ physical Mpc (see Table \ref{tab:galaxy_properties}).
If these size estimates are correct, all three of the closely-separated ($D<2$ physical Mpc) galaxy pairs in our sample (COS-851423 $+$ COS-854905, COS-940214 $+$ COS-1009842, and COS-1009842 $+$ COS-1053257) will each occupy a single ionized bubble.
This is likely a conservative conclusion given that we are ignoring the ionizing photon contribution from fainter ($\Muv{} \gtrsim -20$) sources within our targeted ground-based imaging field. 
The spectroscopic overdensity we have confirmed exists in our observed $z\simeq6.8$ volume ($N / \langle N\rangle \gtrsim 3$) suggests that such faint sources are abundant and hence perhaps contribute substantially to the local ionizing photon budget.
If this is indeed the case, we are likely significantly underestimating the sizes of ionized bubbles around each of our identified UV-bright ($\Muv{} \leq -20.4$) galaxies, possibly to the extent that all their surrounding bubbles have overlapped.
We consider whether the entire 140 physical Mpc$^3$ volume spanned by our confirmed \Lya{} emitters could plausibly have been reionized by considering ionizing photon contribution from all $\Muv{} \leq -17$ galaxies in this region (see also \citealt{RodriguezEspinosa2021}).
Here, we assume an overdensity of $N / \langle N\rangle = 3$ and apply Eq. \ref{eq:bubbleSize} with fixed values of $\langle \xiion{} \rangle = 10^{25.8}$ erg$^{-1}$ Hz and age = 200 Myr.
With these assumptions, we find that our observed overdense field could have been reionized by $\Muv{} \leq -17$ galaxies if their escape fractions are at least $\langle \fesc{} \rangle \geq 6$\%.
Such modest escape fractions are consistent with those measured from star-forming galaxies at $z\lesssim3$ \citep[e.g.][]{Izotov2016a,Izotov2016b,Izotov2018a,Izotov2018b,Shapley2016,Steidel2018,Vanzella2018,Fletcher2019,Pahl2021,Begley2022,Flury2022,Saxena2022}, implying that our targeted region could plausibly be fully reionized.
In this scenario, the 140 physical Mpc$^3$ volume spanned by our confirmed \Lya{} emitters translates to a characteristic bubble radius of $R = (3V/4\pi)^{1/3} = 3.2$ physical Mpc.

\subsection{Enhanced Ly\boldmath{$\alpha$} Emission: A Very Large Ionized Bubble?} \label{sec:bubble}

If our observed galaxies in fact occupy a single very large ($R = 3.2$ physical Mpc) ionized bubble, we would expect their \Lya{} photons to cosmologically redshift far into the damping wing before encountering intergalactic HI and thus transmit efficiently through the IGM.
Below, we explicitly calculate this transmission efficiency and compare to the case when \Lya{} is emitted within a more moderate-sized ($R = 0.7-1.25$ physical Mpc) bubble that might be generated by an individual UV-bright ($-20.5 \leq \Muv{} \leq -22.5$) $z=7$ galaxy (see Eq. \ref{eq:bubbleSize}).
We adopt the damping wing transmission calculations of \citet{MiraldaEscude1998} and assume that the IGM within an ionized bubble has a residual HI fraction of $\xHI{} \lesssim 10^{-5}$ \citep{Mason2020} while the IGM outside the bubble is fully neutral.
We also assume a \Lya{} velocity offset range of 100--500 km s$^{-1}$ as expected for UV-bright $z\simeq7$ systems (\citealt{Maiolino2015,Willott2015,Pentericci2016,Stark2017,Mason2018_IGMneutralFrac,Hashimoto2019,Endsley2022_REBELS}).
With these assumptions, we find that a high fraction ($\approx$80\%) of \Lya{} photons would transmit through the IGM when emitted from a UV-bright $z=7$ galaxy at the center of a very large ($R = 3.2$ physical Mpc) ionized bubble.
When \Lya{} photons are instead emitted from the center of a more moderate-sized ($R = 0.7-1.25$ physical Mpc) bubble, the expected IGM transmission fraction drops significantly to $T\approx30-60$\%, a factor of 1.3--2.6$\times$ lower.

We now explore whether our Binospec data support the presence of a very large ($R\sim3$ physical Mpc) ionized bubble within our observed overdense volume by testing for enhanced \Lya{} emission among our sample.
To do so, we first compare the median \Lya{} EW from the ten UV-bright ($\Muv{} \leq -20.4$) galaxies observed in this work to that from a larger (N=22) Binospec-targeted sample of equally luminous $z\simeq6.8$ galaxies described in \citetalias{Endsley2021_LyA}.
This \citetalias{Endsley2021_LyA} sample was assembled by applying the same Lyman-break selection technique (see \S\ref{sec:observations}) across a much wider area (3 deg$^2$ vs. 0.05 deg$^2$), thereby providing a more cosmologically-averaged census of \Lya{} visibility among UV-bright $z\simeq7$ galaxies.
As described in \citetalias{Endsley2021_LyA}, the median \Lya{} EW of this wider-area sample was inferred to be 10$\pm$3 \AA{}.
In contrast, the ten UV-bright galaxies observed in this work reveal a median \Lya{} EW of 20 \AA{} (see Table \ref{tab:LyA}).
This suggests that the \Lya{} emission of UV-bright ($\Muv{} \leq -20.4$) $z\simeq6.8$ galaxies within our targeted overdense field is enhanced by a factor of $2.0^{\scaleto{+0.9}{4.5pt}}_{\scaleto{-0.5}{4.5pt}}$ relative to average.
As discussed below, this boost in the median \Lya{} EW is unlikely to be driven by abnormal galaxy properties among our sample.

To further test for enhanced \Lya{} visibility within our targeted overdense volume, we now compare the fraction of our galaxies showing strong (EW$>$25 \AA{}) \Lya{} emission against that from other UV-bright $z\sim7$ galaxy samples in the literature.
Using spectroscopic observations covering multiple deep \HST{} fields, \citet{Pentericci2018} find a strong \Lya{} emitter fraction of 9$^{\scaleto{+7}{4.5pt}}_{\scaleto{-4}{4.5pt}}$\% among $\Muv{} < -20.25$ $z\sim7$ galaxies, in agreement with the value of 8$^{\scaleto{+5}{4.5pt}}_{\scaleto{-4}{4.5pt}}$\% from \citet{Schenker2014}.
In contrast, 40\% (4/10) of the similarly luminous ($\Muv{} \leq -20.4$) galaxies observed in this work show \Lya{} EW$>$25 \AA{} (see Table \ref{tab:LyA}).
This factor of 4--5 increase in the strong \Lya{}-emitter fraction is expected if the EWs of our galaxies are typically boosted by a factor of approximately 2, at least assuming the average UV-bright ($\Muv{} \leq -20.4$) $z\simeq7$ \Lya{} EW distribution inferred in \citetalias{Endsley2021_LyA}.
The \citetalias{Endsley2021_LyA} log-normal \Lya{} EW distribution (approximate median EW of 10 \AA{} with 0.35 dex scatter) implies a typical strong \Lya{}-emitter fraction of $\approx$12\%, consistent with the findings of \citet{Pentericci2018} and \citet{Schenker2014}.
If we boost the median EW of this distribution by a factor of 2 (to 20 \AA{}) at fixed scatter, the strong \Lya{}-emitter fraction rises to $\approx$40\% as seen among our observed sample. 

The two metrics of \Lya{} visibility described above (the median EW and the strong \Lya{}-emitter fraction) both indicate that our observed UV-bright ($\Muv{} \leq -20.4$) $z\simeq6.8$ galaxies show significantly enhanced \Lya{} emission.
As motivated in \S\ref{sec:BEAGLE}, the factor of $2.0^{\scaleto{+0.9}{4.5pt}}_{\scaleto{-0.5}{4.5pt}}$ enhancement in median \Lya{} EW among our galaxies relative to the wider-area \citetalias{Endsley2021_LyA} sample is unlikely to be driven by differences in galaxy physical conditions. 
These two samples have nearly equal median hydrogen ionizing photon production efficiencies (\logxiion{} = 25.81 vs. 25.82), V-band dust optical depths ($\tau_{_\mathrm{V}} = 0.02$ vs. 0.03), and \OIIIHb{} EWs (790 vs. 710 \AA{}), implying a very similar efficiency of \Lya{} photon production and escape (\S\ref{sec:BEAGLE}).
The $\approx$2$\times$ increased median \Lya{} EW of our observed galaxies may instead indicate that the large-scale volume (140 physical Mpc$^3$) hosting these systems is highly reionized.
As detailed above, it is feasible that the \Lya{} damping wing transmission of our galaxies would be twice that typical of UV-bright $z\sim7$ systems if our targeted field contains a single very large ($R = 3.2$ physical Mpc) ionized bubble.
The presence of such a large bubble around our galaxies is qualitatively consistent with expectations given the strong galaxy overdensity ($N/\langle N\rangle \gtrsim 3$) we have confirmed exists in our observed volume (e.g. \citealt{Barkana2004,Wyithe2005,Hutter2017,Garaldi2022,Qin2022}).

The \Lya{} profiles of our galaxies offer another way to explore whether our targeted $z\simeq6.8$ volume may contain a very large ($R \gtrsim 3$ physical Mpc) ionized bubble.
Prior to the completion of reionization, \Lya{} photons emitted blueward of systemic velocity will often be completely undetectable given their extremely high optical depth to intergalactic HI.
Hence, blue-sided \Lya{} can only become significantly transparent at $z\gtrsim7$ if the emitting galaxy resides in a large ionized bubble, enabling the photons to cosmologically redshift into the red part of the damping wing before reaching the edge of the bubble \citep[e.g.][]{Matthee2018,Mason2020,Gronke2021,Meyer2021,Smith2021}.
Below, we search for evidence of blue-sided \Lya{} emission from our Binospec-detected galaxies.
We note that, even if our galaxies do occupy a large bubble, we should not necessarily expect them to show blue-sided \Lya{} emission since these photons can also be heavily attenuated by HI within the CGM \citep[e.g.][]{Henry2015,Gazagnes2020}.
Nonetheless, the detection of blue-sided \Lya{} emission from even just one of our galaxies would necessitate the presence of a large HII region within our targeted volume.

We presented the  \Lya{} line 
profiles of our sample in \S\ref{sec:spectra}. 
One of our galaxies (COS-1009842, see \S\ref{sec:spectra} and Fig. \ref{fig:LyA_Detections}) was shown to exhibit a highly symmetric \Lya{} profile. As we motivate above, this may provide additional evidence for a large ionized bubble in our observed field.
Because resonant interactions with neutral hydrogen will distort \Lya{} profiles, one interpretation of the symmetry is that the observed profile closely matches the intrinsic profile, with the observed peak tracing systemic velocity (see e.g. the $z\approx2$ galaxy XLS-20 in \citealt{Naidu2022} for one such known example).
In this scenario, we are detecting \Lya{} photons emitted both blueward and redward of systemic.
The implied scarcity of resonant interactions would necessitate that the observed \Lya{} photons are escaping COS-1009842 through a highly-ionized channel in the ISM and CGM, thus indicating a non-unity HI covering fraction and hence efficient ionizing photon escape from this system (e.g. \citealt{Izotov2018a,Jaskot2019,Gazagnes2020,Naidu2022}).
Furthermore, because \Lya{} photons blueward of systemic velocity are extremely sensitive to neutral hydrogen, the observed symmetry would require a very large, highly-ionized bubble around COS-1009842.
To preserve the line symmetry to within $\approx$10\% across the width of the line (FWHM=260 km s$^{-1}$; see Table \ref{tab:LyA}), we estimate that the \Lya{} photons must travel through a ionized pathlength of at least $\approx$3 physical Mpc along the line of sight (see figure 1 of \citealt{Mason2020}).
However, this is not the only potential interpretation of the symmetric \Lya{} profile observed from COS-1009842. 
We cannot currently rule out the possibility that this source possesses abnormal outflow properties that would lead to a symmetric \Lya{} profile with a peak centered substantially redward of systemic velocity.
Ultimately, a systemic redshift measurement from e.g. [OIII]$\lambda$5007 or [CII]$\lambda$158$\mu$m will be required to determine if the \Lya{} profile of COS-1009842 implies efficient ionizing photon escape as well a surrounding very large ($R\gtrsim3$ physical Mpc) ionized bubble.

\Lya{} spectra of fainter ($\Muv{} > -20$) $z\simeq6.8$ sources within our observed field can further help characterize the surrounding IGM ionization state.
Because fainter $z\sim7$ galaxies are expected to possess lower \Lya{} velocity offsets \citep{Erb2014,Stark2017,Mason2018_IGMneutralFrac}, their photons are intrinsically closer to line center and are thus more susceptible to intergalactic HI if emitted within relatively small ionized bubbles.
Accordingly, if our targeted region of COSMOS indeed contains a very large ($R = 3.2$ physical Mpc) bubble, the \Lya{} EW enhancement among the fainter population should exceed the factor of $\approx$2 increase we find among our UV-bright ($\Muv{} \leq -20.4$) sample.
To quantify this more directly, we consider the scenario that a typical $\Muv{} = -19$ galaxy at $z=7$ has a \Lya{} velocity offset of 100 km s$^{-1}$ \citep{Mason2018_IGMneutralFrac} and lies at the center of a small ionized bubble with $R = 0.4$ physical Mpc (Eq. \ref{eq:bubbleSize})
The expected \Lya{} damping wing transmission of such a source is only $T=15$\%, a factor of $>$5$\times$ lower than if the galaxy were to lie at the center of a very large ($R > 3$ physical Mpc) ionized bubble ($T \gtrsim 80$\%).
If $\gtrsim$80\% of \Lya{} photons emitted by faint galaxies in our targeted $z\simeq6.8$ volume do in fact transmit through the IGM, we would expect to see a small evolution in the \Lya{}-emitter fraction from $z\sim6$ \citep{Castellano2018}, in contrast to what is generally reported in the literature \citep[e.g.][]{Stark2011,Caruana2014,Schenker2014,Tilvi2014,deBarros2017,Pentericci2018}.

Because the wide-area COSMOS field is currently only covered by ground-based near-infrared imaging, we cannot yet identify faint galaxies in our targeted $z\simeq6.8$ volume for \Lya{} follow-up.
However, upcoming \JWST{} observations with the COSMOS-Webb program will partially cover this field (see Fig. \ref{fig:overdensity}) with 5$\sigma$ depths of $m\approx27.5-28$ in the NIRCam F115W, F150W, F277W, and F444W filters \citep{COSMOSWebbProposal}.
Adopting the $z\sim7$ luminosity function from \citealt{Bouwens2021_LF}, we estimate that COSMOS-Webb observations will enable the detection of $\gtrsim$100 $z=6.7-6.9$ galaxies with $\Muv{} \leq -19$ ($m<28$) within our targeted field assuming an overdensity of $N/\langle N\rangle \gtrsim 3$.
Deep \Lya{} follow-up observations of these systems would clearly help distinguish whether our targeted region of COSMOS contains many small-to-moderate sized ($R \sim 0.5-1$ physical Mpc) ionized bubbles or a single very large ($R \gtrsim 3$ physical Mpc) bubble.

\section{Summary and Outlook} \label{sec:summary}

In this work, we have sought to better characterize the size and associated overdensity of a tentative large ($R>1$ physical Mpc) ionized bubble at $z\simeq6.8$ we previously identified in \citetalias{Endsley2021_LyA}.
To accomplish this goal, we conducted deep \Lya{} spectroscopy of ten UV-bright ($\Muv{} \leq -20.4$) Lyman-break galaxies at $z\simeq6.6-6.9$ surrounding this possible ionized bubble in an 11$\times$15 arcmin$^2$ area.
Below is a summary of our results and conclusions:
\begin{enumerate}

    \item We confidently detect (S/N$>$7) \Lya{} emission in nine of ten targeted galaxies, yielding redshifts spanning \zLya{} = 6.701--6.882 consistent with our photometric selection window. The \Lya{} fluxes range between (3.7--18.5)$\times$10$^{-18}$ erg/s/cm$^2$ and line widths span FWHM$\approx$150--540 km s$^{-1}$ with a median of 272 km s$^{-1}$.
    
    \item We quantify the spectroscopic overdensity within the 140 physical Mpc$^3$ volume spanned by our \Lya{}-detected sources. This volume contains at least $3.0^{\scaleto{+9.0}{4.5pt}}_{\scaleto{-0.9}{4.5pt}}$ times the number of $\Muv{} \leq -21$ systems expected on average from wide-area luminosity functions \citep{Bowler2017,Harikane2021}. Our spectra therefore confirm that this region of COSMOS traces a strong, large-scale galaxy overdensity at $z=6.8$. 
    
    \item We estimate that each of our UV-bright ($\Muv{} \leq -20.4$) galaxies could potentially generate ionized bubbles with radii spanning $R = 0.69-1.13$ physical Mpc. 
    These size estimates suggest that the three closely-separated ($D = 0.2-1.8$ physical Mpc) galaxy pairs in our sample may each occupy a single ionized bubble due to overlap. 
    If fainter ($\Muv{} \gtrsim -20$) galaxies contribute significantly to the ionizing photon budget, the entire overdense volume spanned by our confirmed \Lya{} emitters (characteristic radius $R = 3.2$ physical Mpc) may potentially be reionized.
    In this scenario, we would expect a typical \Lya{} damping wing transmission fraction of $T\approx80$\% among our observed UV-bright galaxies.
    If these systems instead occupy more moderate-sized ($R \sim 1$ physical Mpc) bubbles, their \Lya{} IGM transmission fraction would be $\approx$2$\times$ lower ($T\approx30-60$\%).
    
    \item We test for enhanced \Lya{} emission among our sample to explore whether a very large ($R \sim 3$ physical Mpc) ionized bubble may exist within our targeted overdense volume.
    We detect strong \Lya{} emission (EW$>$25 \AA{}) from 40\% (4/10) of our observed UV-bright galaxies, a factor $\approx$4--5 higher than that typically found among similarly luminous ($\Muv{} < -20.25$) $z\sim7$ systems \citep{Schenker2014,Pentericci2018}. 
    The median \Lya{} EW of our targeted sample ($\geq$20 \AA{}) is also enhanced by at least a factor of $2.0^{\scaleto{+0.9}{4.5pt}}_{\scaleto{-0.5}{4.5pt}}$ relative to that inferred from a larger (N=22) sample of equally luminous $z\simeq6.6-6.9$ galaxies (EW=10$\pm$3 \AA{}; \citetalias{Endsley2021_LyA}). 
    We find no clear evidence of significant differences in the typical \Lya{} photon production or escape efficiency between our sample and that of \citetalias{Endsley2021_LyA} based on the inferred galaxy properties from SED fitting.
    The enhanced \Lya{} emission of our sample may instead indicate that the 140 physical Mpc$^3$ volume hosting these systems is highly reionized, a scenario that is qualitatively consistent with expectations given the surrounding strong, large-scale galaxy overdensity \citep[e.g.][]{Barkana2004,Hutter2017,Qin2022}.
    
    \item We investigate whether the \Lya{} profiles from our sample also support the presence of a surrounding very large ionized bubble \citep[e.g.][]{Matthee2018}.
    One of our galaxies (COS-1009842; \zLya{} = 6.76) displays a \Lya{} velocity profile that is indistinguishable from a symmetric Gaussian, potentially suggesting that the peak of this line traces systemic velocity.
    If this scenario is confirmed, it would indicate that blue-sided \Lya{} photons emitted from COS-1009842 face very little optical depth from the surrounding IGM (as well as from the ISM and CGM), thereby necessitating a very large ($R \gtrsim 3$ physical Mpc) ionized bubble within our observed volume \citep{Mason2020}. 

    \item \Lya{} spectra of fainter ($\Muv{} > -20$) $z\simeq6.8$ galaxies can further help characterize the IGM ionization state within our observed overdense field. 
    Given the expected smaller velocity offsets of fainter sources \citep{Erb2014,Stark2017,Mason2018_IGMneutralFrac}, such galaxies may show up to a factor of $\sim$5 increase in \Lya{} EW relative to average if they occupy a very large ($R > 3$ physical Mpc) ionized bubble. 
    We estimate that the \JWST{} COSMOS-Webb survey will detect $\gtrsim$100 faint ($\Muv{} \leq -19$) $z=6.7-6.9$ galaxies within our targeted field assuming an overdensity of $N/\langle N\rangle \gtrsim 3$. 
    Deep \Lya{} follow-up observations of these fainter systems would clearly help distinguish whether our targeted region of COSMOS contains many small-to-moderate sized ($R\sim0.5-1$ physical Mpc) ionized bubbles or a single very large ($R\gtrsim3$ physical Mpc) bubble.
    
\end{enumerate}

This study demonstrates how wide-area $z\gtrsim7$ \Lya{} spectroscopy can help identify and characterize large ionized bubbles formed during reionization.
Upcoming facilities such as the Nancy Grace Roman Space Telescope as well as the Giant Magellan Telescope will enable the selection and spectroscopic follow-up of massive, UV-luminous $z\gtrsim7$ galaxies across much wider areas (1000 deg$^2$), allowing the opportunity to further study the largest HII regions present in the early Universe.
Combining such studies with 21cm observations from e.g. LOFAR, HERA, and SKA will deliver rich insight into how intergalactic HII regions formed and grew during reionization.

\section*{Acknowledgements}

The authors sincerely thank the anonymous referee for their helpful, constructive comments. The authors also thank John Chisholm for insightful conversations which benefited this work, as well as St{\'e}phane Charlot and Jacopo Chevallard for providing access to the \textsc{beagle} SED fitting code.
Both authors acknowledge funding from JWST/NIRCam contract to the University of Arizona, NAS5-02015.
DPS acknowledges support from the National Science Foundation through the grant AST-2109066.

Observations reported here were obtained at the MMT Observatory, a joint facility of the University of Arizona and the Smithsonian Institution.
RE sincerely thanks the MMT queue observers Michael Calkins, Ryan Howie, ShiAnne Kattner, and Skyler Self for their assistance in collecting the Binospec data, as well as Ben Weiner for managing the queue.
Based on data products from observations made with ESO Telescopes at the La Silla Paranal Observatory under ESO programme ID 179.A-2005 and on data products produced by CALET and the Cambridge Astronomy Survey Unit on behalf of the UltraVISTA consortium.
This research has made use of the NASA/IPAC Infrared Science Archive, which is funded by the National Aeronautics and Space Administration and operated by the California Institute of Technology.
This work is based [in part] on observations made with the Spitzer Space Telescope, which was operated by the Jet Propulsion Laboratory, California Institute of Technology under a contract with NASA. 

The Hyper Suprime-Cam (HSC) collaboration includes the astronomical communities of Japan and Taiwan, and Princeton University. The HSC instrumentation and software were developed by the National Astronomical Observatory of Japan (NAOJ), the Kavli Institute for the Physics and Mathematics of the Universe (Kavli IPMU), the University of Tokyo, the High Energy Accelerator Research Organization (KEK), the Academia Sinica Institute for Astronomy and Astrophysics in Taiwan (ASIAA), and Princeton University. Funding was contributed by the FIRST program from the Japanese Cabinet Office, the Ministry of Education, Culture, Sports, Science and Technology (MEXT), the Japan Society for the Promotion of Science (JSPS), Japan Science and Technology Agency (JST), the Toray Science Foundation, NAOJ, Kavli IPMU, KEK, ASIAA, and Princeton University. 
This paper makes use of software developed for Vera C. Rubin Observatory. We thank the Rubin Observatory for making their code available as free software at http://pipelines.lsst.io/.
This paper is based on data collected at the Subaru Telescope and retrieved from the HSC data archive system, which is operated by the Subaru Telescope and Astronomy Data Center (ADC) at NAOJ. Data analysis was in part carried out with the cooperation of Center for Computational Astrophysics (CfCA), NAOJ. We are honored and grateful for the opportunity of observing the Universe from Maunakea, which has the cultural, historical and natural significance in Hawaii. 

This research made use of \textsc{astropy}, a community-developed core \textsc{python} package for Astronomy \citep{astropy:2013, astropy:2018}; \textsc{matplotlib} \citep{Hunter2007_matplotlib}; \textsc{numpy} \citep{harris2020_numpy}; and \textsc{scipy} \citep{Virtanen2020_SciPy}.

\section*{Data Availability}

The optical through mid-infrared imaging data underlying this article are available through their respective data repositories. See \url{https://hsc-release.mtk.nao.ac.jp/doc/} for HSCSSP data, \url{https://hsc-release.mtk.nao.ac.jp/doc/index.php/chorus/} for CHORUS data, \url{https://irsa.ipac.caltech.edu/data/COSMOS/} for \HST{} F814W data, \url{http://www.eso.org/rm/publicAccess#/dataReleases} for UltraVISTA data, and \url{https://sha.ipac.caltech.edu/applications/Spitzer/SHA/} for IRAC data. The MMT/Binospec data will be shared upon reasonable request to the corresponding author.

%%%%%%%%%%%%%%%%%%%%%%%%%%%%%%%%%%%%%%%%%%%%%%%%%%



\bibliographystyle{mnras}
\bibliography{paper_ref} % if your bibtex file is called example.bib

%%%%%%%%%%%%%%%%%%%%%%%%%%%%%%%%%%%%%%%%%%%%%%%%%%

%%%%%%%%%%%%%%%%% APPENDICES %%%%%%%%%%%%%%%%%%%%%

\appendix
 
%%%%%%%%%%%%%%%%%%%%%%%%%%%%%%%%%%%%%%%%%%%%%%%%%%

% Don't change these lines
\bsp	% typesetting comment
\label{lastpage}
\end{document}